\documentclass[12pt]{article}
\usepackage{fullpage}
\usepackage{amsmath}
\usepackage{amssymb}
\usepackage{epsfig}
\usepackage{fullpage}
\usepackage{subfigure}
\usepackage{feynmf}
\begin{document}


\begin{center}
{\Large \bf  Rare charm meson decays $D\to Pl^+l^-$  \\
 and $c \to ul^+ l^-$ in SM and MSSM}

\vspace{1cm}
{\large \bf S. Fajfer$^{a}$,  S. Prelovsek$^{a}$ and
P. Singer$^{b}$ \\}
\vspace{.5cm}

{\it a) Department of Physics, University of Ljubljana, Jadranska 19, 1000 Ljubljana,
Slovenia and

J. Stefan Institute, Jamova 39,  1000 Ljubljana, Slovenia}

 \vspace{.3cm}

{\it b) Department of Physics, Technion - Israel Institute  of Technology,
Haifa 32000, Israel}

\end{center}

\vspace{0.25cm}

\centerline{\large \bf ABSTRACT}

\vspace{0.5cm}

We study the nine possible rare charm meson decays $D\to Pl^+l^-$ ($P=\pi,K,\eta,\eta^\prime$) using the Heavy Meson Chiral Lagrangians and find them to be dominated by the long distance contributions.  The decay $D^+\to \pi^+l^+l^-$ with the branching ratio $\sim 1\times 10^{-6}$ is expected to have the best chances for an early  experimental discovery. The short distance contribution in the five Cabibbo suppressed channels arises via the  $c\to ul^+l^-$ transition; we find that this contribution is detectable only in the  $D\to \pi l^+l^-$ decay, where
it dominates the differential spectrum at high-$q^2$. The general Minimal Supersymmetric Standard Model can enhance the $c\to ul^+l^-$ rate by up to an order of magnitude; its effect on the $D\to Pl^+l^-$ rates is small since the $c\to ul^+l^-$ enhancement is sizable in low-$q^2$ region, which is inhibited in the hadronic decay. 

\vspace{1cm}

\section{Introduction}

The flavour-changing neutral processes are rare in
the standard model and  are of obvious 
interest in the search for new physics.  
Processes  like $c\to u\gamma$ and $c\to ul^+l^-$  are screened by 
the long distance contributions in the decays of charm hadrons  \cite{LD,FPS.vll} and one has to look for
specific
hadronic observables \cite{FPS.bc,FPSW,special} in order to probe possible new physics
\cite{sasa,masiero.cugamma}. The long distance contributions are
expected
to dominate over the short distance contributions also in the $D^0- \bar
D^0$
mixing \cite{nelson}, for which interesting experimental results have been
reported recently \cite{mixing}.

The  long distance and the short 
distance
contributions
to rare charm meson decays $D\to Vl^+l^-$ with  $V=\rho,\omega,\phi,K^*$
have
been considered in  \cite{FPS.vll}. The long distance contributions were
shown
to be largely dominant and screen  possible effects of new physics in $c\to ul^+l^-$, unless
these are very large. The
experimental upper bounds on their branching ratios are presently 
in the $10^{-5}$ range
\cite{vll.exp} and are an order of magnitude larger than the standard model
prediction for specific channels \cite{FPS.vll}. The decay $D_s^+\to
\rho^+l^+l^-$ is predicted at the highest rate $\sim 3\times 10^{-5}$
\cite{FPS.vll}, but there is unfortunately no experimental data on this
particular channel.

In the present paper we consider  the  weak decays $D\to Pl^+l^-$ with
pseudoscalar
$P=\pi, K,\eta,\eta^\prime$, some of which having contribution from  the  $c\to ul^+l^-$
transition.
These channels have not been observed so far and only  experimental upper
bounds on the various branching ratios in the range $10^{-6}-10^{-4}$ exist
\cite{pll.exp,pll.focus,PDG}. The  recent E791 analysis \cite{pll.exp}
considers
all $D^+$ and $D^+_s$ decay channels. The most recent FOCUS analysis
\cite{pll.focus}
provides
upper bounds of about $8\times 10^{-6}$ on the $D^+\to \pi^+\mu^+\mu^-$ and
$D^+\to K^+\mu^+\mu^-$ branching ratios and is not far from our  standard model prediction $ 1\times 10^{-6}$ for $D^+\to
\pi^+\mu^+\mu^-$. 
The limits on  $D^0$ and $D^+$ modes at the level $10^{-6}$ are expected from CLEO-c and B-factories, while the limits on $D_s^+$ modes are expected to be an order of magnitude  milder \cite{CLEOc}.

On the theoretical side, the  long distance  contributions to
$D \to
\pi l^+l^-$ decays have been considered in \cite{singer}. We consider here
also
the  long distance weak annihilation contribution  and confirm it
to be
small in this  channel.
Calculations for other $D \to P l^+ l^-$ channels are not available in the
literature. In the present work we investigate all these channels, including
long-distance (LD) and possible short-distance (SD) contributions
arising from the $c \to u l^+ l^-$  transition.
The QCD corrections to $c\to ul^+l^-$ amplitude have not been studied in
detail
yet and we  incorporate only what we believe to be the most
important QCD effects. We explore also the sensitivity of $c\to ul^+l^-$ transition to  (i) minimal
supersymmetric model with general soft-breaking terms and (ii) two Higgs
doublet
model with flavour changing neutral Higgs interactions.

\vspace{0.2cm}

The $c\to ul^+l^-$ transition in SM, MSSM and Two Higgs Doublet model
is studied in Section 2. The long distance contributions are considered within the
Heavy Meson Chiral Lagrangian approach in Section 3. The results are
compiled in
Section 4, while conclusions are given in Section 5.

\section{The $c\to ul^+l^-$ decay}

The Lagrangian leading to $c\to ul^+l^-$ transition is (using notation as in
\cite{lunghi})
\begin{align}
\label{ope.lunghi}
{\cal L}=-\frac{4G_F}{\sqrt{2}}V_{cs}^* V_{us}\bigl[ c_7 {\cal O}_7+
c_7^\prime {\cal
O}_7^\prime+\frac{\alpha}{4\pi}\bigl\{c_9 {\cal O}_9+ c_9^\prime {\cal
O}_9^\prime+c_{10} {\cal O}_{10}+ c_{10}^\prime {\cal
O}_{10}^\prime \bigr\}\bigr]
\end{align}
where
\begin{alignat}{3}
{\cal O}_7&=\frac{e}{16\pi^2}m_c \bar u\sigma^{\mu\nu}P_Rc~F_{\mu\nu}&\qquad
{\cal O}_9&= \bar u\gamma^\mu P_Lc~\bar l\gamma_\mu l&\qquad
{\cal O}_{10}&= \bar u\gamma^\mu P_Lc~\bar l\gamma_\mu\gamma_5 l\\
{\cal O}_7^\prime&=\frac{e}{16\pi^2}m_c \bar
u\sigma^{\mu\nu}P_Lc~F_{\mu\nu}&\quad {\cal O}_9^\prime&= \bar u\gamma^\mu
P_Rc~\bar l\gamma_\mu l&\quad {\cal O}_{10}^\prime&= \bar u\gamma^\mu
P_Rc~\bar
l\gamma_\mu\gamma_5 l \nonumber
\end{alignat}
with $P_{R,L}=(1\pm \gamma_5)/2$. In Eq. (\ref{ope.lunghi}) only the CKM matrix element $V^*_{cs} V_{us}$ appears, for reasons
explained
in the next subsection \cite{GHMW}.
The Wilson coefficients in  various scenarios are given in the following sections. The differential branching ratio is
given by
\cite{lunghi}
\begin{align}
\label{br}
&\frac{dBr(c\to ul^+l^-)}{ds}\equiv \frac{1}{\Gamma(D^0)}~\frac{d\Gamma(c\to u l^+l^-)}{ds}=\biggl[{G_F^2m_c^5\over
192\pi^3
\Gamma(D^0)}\biggr]\frac{\alpha^2}{4\pi^2}|V_{cs}^*V_{us}|^2(1-s)^2\nonumber\\
&\times
\biggl[\bigl\{(1+2s)(|c_9|^2+|c_{10}|^2)+4(1+2/s)|c_7|^2+12Re[c_7^*c_9]\bigr\}+
\bigl\{c_{7,9,10}\to c_{7,9,10}^\prime\bigr\}\biggr]
\end{align}
where $s=m_{ll}^2/m_c^2$, $m_c\simeq 1.5$ GeV and the mass of $l=e,\mu$ is neglected. The
short-distance part of the $D\to Pl^+l^-$ amplitude, which is induced by
$c\to
ul^+l^-$ transition, is given by (\ref{asd}) in Appendix.

\subsection{Standard model}

The $c\to ul^+l^-$ amplitude is given by the  $\gamma$ and $Z$ penguin
diagrams
and $W$ box diagram at one-loop electroweak order in the standard model,
and is dominated by the light quarks in the loop. One has \cite{FPS.vll,il}
\begin{equation}
\label{c0.sm}
c_9(m_W) \simeq\frac{4}{9}\ln\frac{m_s}{m_d}=
1.34\pm 0.09~,\quad c_{7,10}(m_W)\propto\frac{m_{d,s}^2}{m_W^2}\simeq 0~, \quad c^\prime_{7,9,10}(m_W)\propto
\frac{m_u}{m_c}c_{7,9,10}\simeq 0~
\end{equation}
for $m_s/m_d=21\pm 4$ MeV \cite{PDG}, where the terms proportional to $m_{d,s}^2/m_W^2$ have been  neglected. The leading  term $\ln(m_s/m_d)$ in $c_9$ 
 arises from the  penguin diagram with photon  emitted
from the intermediate quark.

\begin{figure}[!htb]
\begin{center}
\includegraphics[scale=0.5]{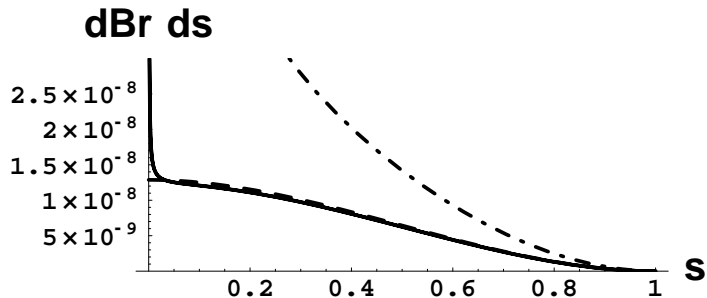}
 \caption{The differential branching ratio $dBr(c\to ul^+l^-)/ds$: the dashed line denotes the one-loop standard model prediction, while the solid line incorporates also the QCD corrections to $c_7$ \cite{GHMW}. The  best enhancement of the $c\to ul^+l^-$ rate in the general MSSM  is given by the
dot-dashed line, where the mass insertions are taken  at their maximal
values
(\ref{insertionlr}, \ref{insertionll}) and $\alpha_s=0.12$, $M_{sq}=M_{gl}=250$ GeV.  }
\label{fig1}
\end{center}
\end{figure}

The QCD corrections to $c\to ul^+l^-$ amplitude have not been studied in
detail yet.  The QCD corrections to $c_7$, which is extremely small at the one-loop level, have been studied in  \cite{GHMW} and are found to be large 
\begin{equation}
\label{two.loop}
c_7^{eff}(m_c)=-(0.007+0.020 ~i)[1\pm 0.2]~.
\end{equation}
We expect the QCD corrections to $c_9$  to be rather unimportant,  given that $c_9$ is relatively large already at one-loop level \cite{buras}. We assume that the QCD corrections to $c_{10}$ do not affect the $c\to ul^+l^-$ rate significantly and   use therefore only the $c_7$ and $c_9$ coefficients.
The differential branching ratios for the cases with and without QCD
corrections
are shown by solid and dashed lines in Fig. \ref{fig1}, 
respectively. The
 branching ratio $[6\pm 1]\times 10^{-9}$ is small and arises mainly from $c_9$; the contribution from $c_7$ is small in spite of QCD enhancement. 

\subsection{ Minimal supersymmetric standard model }

New sources of flavour violation are present in the minimal supersymmetric
standard
model (MSSM) and these  depend crucially on the mechanism of the
supersymmetry
breaking. The schemes with flavour-universal soft-breaking terms lead to
contributions proportional to $\sum_{q=d,s,b}V_{cq}^*V_{uq}m_q^2$ and have
negligible effect on the $c\to ul^+l^-$ rate \cite{duncan.wyler}. Our
purpose
here is to explore the largest possible enhancement of the $c\to ul^+l^-$
rate
in  general MSSM with {\bf non-universal soft breaking terms}. Based on the experience  from the $c\to
u\gamma$ decay \cite{sasa,masiero.cugamma},    where the dominant contribution
arises from gluino diagrams with the  squark-mass insertion $(\delta^u_{12})_{LR}$, we
concentrate
only on the gluino exchange diagrams with single mass insertion\footnote{We work in
the
super-CKM  basis for squarks, where the squark - quark - gaugino vertex
has the
same flavour structure as the quark - quark - gauge boson vertex; for
review see  \cite{HF}.}. Following the
analogous calculation for $b\to sl^+l^-$ \cite{lunghi}, we get for the Willson coefficients in the MSSM
\begin{align}
\label{c.susy}
c_7^{gluino}&=\frac{e_u}{e_d}\frac{\sqrt{2}}{M_{sq}^2G_F}\frac{1}{3}\frac{N_c^2-1}{2N_c}\frac{\pi\alpha_s}{V_{cs}^*V_{us}}
\bigl[(\delta_{12}^u)_{LL}\frac{1}{4}P_{132}(z)+(\delta_{12}^u)_{RL}P_{122}(z)\frac{M_{gl}}{m_c}\bigr]\leq
0.2\\
\label{c.susy1}
c_9^{gluino}&=-\frac{e_u}{e_d}\frac{\sqrt{2}}{M_{sq}^2G_F}\frac{1}{3}\frac{N_c^2-1}{2N_c}\frac{\pi\alpha_s}{V_{cs}^*V_{us}}
\frac{1}{3}P_{042}(z)(\delta_{12}^u)_{LL}\leq 0.002\\
c_{10}^{gluino}&\simeq 0
\end{align}
with  $\alpha_s\!=\!\alpha_s(m_W)=0.12$, $N_c\!=\!3$, $z=M_{gl}^2/M_{sq}^2$, $P_{ijk}(z)=\int_0^1
dx\int_0^1
dy~ y^i(1-y)^j[1-y+zxy+z(1-x)y]^{-k}$, $e_u=2/3$ and $e_d=-1/3$. The numerical bounds in (\ref{c.susy},\ref{c.susy1})
 are obtained by using parameter values as discussed below.
The expressions for $c_{7,9,10}^\prime$ are obtained by replacing
$L\leftrightarrow R$ in the formulas above. 
We  use gluino mass $M_{gl}\!=\! 250$ GeV and the common value for squark masses
$M_{sq}\!=\!250$ GeV,  given by the lower experimental bounds \cite{PDG}.

The mass insertions are free parameters in a general MSSM. The strongest
upper
bound on  $(\delta_{12}^u)_{LR}$ comes by requiring that the minima of the
scalar potential do not break charge or color, and that they are bounded
from
bellow \cite{CD,sasa}, giving
\begin{equation}
\label{insertionlr}
 |\delta_{12}^u|_{LR}~,~|\delta_{12}^u|_{RL}\leq 0.0046~\qquad{\rm for}\quad
M_{sq}=250\ {\rm GeV}.
\end{equation}
The insertions $(\delta_{12}^u)_{LL}$ and $(\delta_{12}^u)_{RR}$ can be
bounded
by saturating the experimental upper bound $\Delta m_D<4.5\times 10^{-14}$
GeV
\cite{mixing} by the gluino exchange \cite{sasa,gabbiani}; the
corresponding constraint  on $(\delta_{12}^u)_{LR}$ is weaker than
(\ref{insertionlr}). Since we are interested in exhibiting the largest
possible  enhancement of the
$c\to
ul^+l^-$ rate, we saturate $\Delta m_D$ by $(\delta_{12}^u)_{LL}$, obtaining
\cite{sasa,gabbiani}
\begin{equation}
\label{insertionll}
 |\delta_{12}^u|_{LL}~\leq 0.03\qquad{\rm for}\quad
M_{sq}=M_{gl}=250\ {\rm GeV}
\end{equation}
and set $(\delta_{12}^u)_{RR}=0$.

The biggest possible enhancement of the $c\to ul^+l^-$ rate is obtained using the  mass insertions at their upper bounds and is shown by the
dot-dashed line in Fig. \ref{fig1}. The effect is dominated by the gluino exchange diagrams induced by $(\delta_{12}^u)_{LR}$  and can enhance the $c\to ul^+l^-$ rate by nearly an  order of magnitude, with the best enhancement displayed in Table \ref{tab1}. 

The supersymmetric enhancement of $c\to ul^+l^-$ is due to the increase in $c_7$ (Eq. \ref{c.susy}) and is manifested at small $m_{ll}$ due to the exchange of an almost real photon. 
This enhancing mechanism is unfortunately not
 present in
 $D\to Pl^+l^-$ decays (see Eq. (\ref{asd})) since the decay $D\to P\gamma$
with
 the real photon in the final state is forbidden (see Eq. (\ref{1})).

\begin{table}[h]
\begin{center}
\begin{tabular}{|c|c|c|}
\hline
&${Br^{SM}}$&${Br^{MSSM}}$\\
&           &best~ enhanc.\\
\hline
$c\to u e^+e^-$&$(6\pm 1)\times 10^{-9}$&$6\times 10^{-8}$\\
$c\to u\mu^+\mu^-$&$(6\pm 1)\times 10^{-9}$&$2\times 10^{-8}$\\
\hline
\end{tabular}
\caption{ 
The second  column
represents
the standard model prediction for $c\to ul^+l^-$ branching ratios, which is practically unaffected by the  QCD corrections (see
text). The third column represents the biggest possible enhancement of the branching ratio in MSSM, evaluated for mass insertions at 
their
maximal
values (\ref{insertionlr}, \ref{insertionll}). 
}
\label{tab1}
\end{center}
\end{table}

\subsection{Flavour changing neutral Higgs}

The
 tree-level exchange of flavour changing neutral Higgs \cite{sher} turns out
to have  a negligible
effect
on $c\to ul^+l^-$ rate, due to the strong constraint coming from
the experimental
upper bound on  $\Delta m_D$ and due to the small mass of the leptons $e$ and $\mu$. Assuming the same $c-u-H$ coupling\footnote{The coupling is $f_{cu}$ for $c-u-H_{1,2}^0$ and $f_{cu}\gamma_5$ for $c-u-A^0$.}  $f_{cu}$ and mass $m_H=300$ GeV for all three  neutral physical Higgses in the Two Higgs Doublet Model, and 
saturating the  experimental upper bound   $\Delta m_D\leq
4.5\times
10^{-14}$ GeV \cite{mixing}\footnote{The matrix elements of four-fermion operators is evaluated according to \cite{gabbiani}.}
\begin{equation}
\frac{4}{3} \frac{f_{cu}^2}{m_{H}^2}f_D^2m_D\leq (\Delta m_D)_{exp}~,
\end{equation}
we get $f_{cu}\leq 2\times 10^{-4}$. This leads to a branching ratio
\begin{equation}
Br(c\to u\mu^+\mu^-)^{H^0}={5m_c^5\over
768\pi^3\Gamma(D^0)}\biggl({f_{cu}m_{\mu}\over vm_{H}^2}\biggr)^2\lesssim  7\times 10^{-16}~.
\end{equation}
Thus,  unlike in the supersymmetric model, the experimental upper bound on
$\Delta m_D$
imposes this  new contribution to be negligible.

The  authors of \cite{CMM} have studied the constraints on the parameters of this model imposed by the present data on the semileptonic and leptonic $D$ decays. Since they did not consider the constraint coming from the $D^0-\bar D^0$ mixing, they have obtained rather mild constraints.

\section{Long distance contributions}

Now we  turn to an estimate of the long distance contributions to
the $D\to P l^+ l^-$ decays. 
The dominant long distance contributions arise via the weak transition $D\to
P\gamma^*$
followed by $\gamma^*\to l^+l^-$. The general Lorentz structure of the $D\to
P\gamma^*$ amplitude, consistent with electro-magnetic gauge invariance,
is \cite{epr}
\begin{equation}
\label{1}
 {\cal A}[D(p)\to P(p^\prime)\gamma^*(q,\epsilon)]\propto
A(q^2)~\epsilon^*_\mu~[q^2(p+p^\prime)^\mu-(m_D^2-m_P^2)q^\mu]~
\end{equation}
and this  amplitude vanishes for the case of a real photon. The factor
$q^2$ in
(\ref{1}) cancels the photon propagator $1/q^2$ and the general amplitude has the form
\begin{equation}
\label{2}
{\cal A}[D(p)\to P\gamma^*\to
Pl^+(p_+)l^-(p_-)]=i~\tfrac{G_F}{\sqrt{2}}~e^2~A(q^2)~\bar u(p_-) \!{\not
\!p}
v(p_+).
\end{equation}

The long distance contribution is induced by the effective nonleptonic weak
Lagrangian
\begin{equation}
\label{eff}
{\cal L}^{|\Delta
c|=1}\!=\!-\tfrac{G_F}{\sqrt{2}}V_{cq_j}^*V_{uq_i}\bigl[a_1
~\bar u\gamma^{\mu}(1-\gamma_5)q_i~\bar q_j\gamma_{\mu}(1-\gamma_5)c
+a_2~\bar q_j\gamma_{\mu}(1-\gamma_5)q_i~\bar
u\gamma^{\mu}(1-\gamma_5)c\bigr]~,
\end{equation}
accompanied by the emission of the virtual photon.
Here $q_{i,j}$ denote the $d$ or $s$ quark fields. The coefficients
$a_1=1.2$
and $a_2=-0.5$ have been determined from the experimental data on
nonleptonic
charm meson decays in the extensive analysis  based on the factorization
approximation of \cite{factor}. We also
systematically undertake the factorization
approximation to evaluate the matrix element for the product of the currents
(\ref{eff}).

In order to treat the transition among physical particles, we shall use
the effective Lagrangian approach with heavy  pseudoscalar $D$, heavy
vector $D^*$, light pseudoscalar $P$ and including also  light vector $V$
degrees of
freedom. The later are necessary, since
they play a dynamical role  in the photon emission from a
meson via vector meson dominance (VMD)  and lead to the resonant spectrum in
terms of
invariant di-lepton mass $m_{ll}$. We organize various effective
interactions
among the mesonic degrees of freedom  following the Heavy Meson Chiral
Lagrangian approach \cite{wise}, which is reviewed
in \cite{casalbuoni} and is most
likely the best suited framework for treating the problem
under investigation.  It embodies two
important global symmetries of QCD: the heavy quark spin and flavour
symmetry
$SU(2N_f)$ in the limit $m_{c}\to \infty$ and chiral symmetry $SU(3)_L\times
SU(3)_R$, spontaneously broken to $SU(3)_V$,  in the limit $m_{u,d,s}\to
0$. The
light vector mesons are introduced  by promoting the symmetry
$G=[SU(3)_L\times
SU(3)_R]_{global}/[SU(3)_V]_{global}$ to  $G^\prime =[SU(3)_L\times
SU(3)_R]_{global}\times[SU(3)_V]_{local}$, where the light vector
resonances are
identified with the gauge bosons of   $[SU(3)_V]_{local}$ \cite{hidden}.
One is
free to fix the gauge of $[SU(3)_V]_{local}$ and the two theories, based
on the
groups $G$ and $G^\prime$, are equivalent up to terms with derivatives
on
the light vector fields  \cite{hidden}.

Keeping only the kinetic and interaction terms of the
lowest
non-trivial order, the Lagrangian has the form  \cite{casalbuoni,thesis}
\begin{align}
\label{lag}
{\cal L}&=  -\tfrac{f^2}{2}\Bigl\{tr[{\cal A}_{\mu}{\cal A}^\mu]+a~tr[({\cal
V}_\mu-\rho_\mu)^2]\Bigl\}+\tfrac{1}{2\tilde
g_V^2}tr[F_{\mu\nu}(\rho)F^{\mu\nu}(\rho)]\nonumber\\
&~\nonumber\\
&+iTr[H_bv_\mu\{\delta_{ba}\partial^\mu-i\tfrac{2}{3}e \delta_{ba}A^\mu+{\cal
V}^\mu_{ba}-\kappa({\cal V}^\mu-\rho^\mu)_{ba}\}\bar H_a]+igTr[H_b\gamma_\mu\gamma_5{\cal
A}^\mu_{ba}\bar H_a]
~,
\end{align}
with
\begin{align*}
{\cal A}_\mu&=\tfrac{1}{2}[\xi^\dagger (\partial_\mu+ie{ Q}
A_\mu)\xi-\xi
(\partial_\mu+ie{ Q} A_\mu)\xi^\dagger]~,\\
 {\cal V}_\mu&=\tfrac{1}{2}[\xi^\dagger (\partial_\mu+ie{ Q}
A_\mu)\xi+\xi
(\partial_\mu+ie{ Q} A_\mu)\xi^\dagger]~,
\end{align*}
  ${ Q}={\rm diag}(2/3,-1/3,-1/3)$ and photon field
$A_\mu$. The light fields are incorporated in
\begin{align}
\xi&=exp~\frac{i}{f}
\begin{pmatrix}
\tfrac{\pi^0}{\sqrt{2}}+[\tfrac{\eta_8}{\sqrt{6}}+
\tfrac{\eta_0}{\sqrt{3}}]&\pi^+ & K^+\\
\pi^-&-\tfrac{\pi^0}{\sqrt{2}}+[\tfrac{\eta_8}{\sqrt{6}}+
\tfrac{\eta_0}{\sqrt{3}}]
&K^0\\ K^-&\bar K^0
&[-\tfrac{2\eta_8}{\sqrt{6}}+\tfrac{\eta_0}{\sqrt{3}}]
\end{pmatrix}~
, \nonumber\\
\rho_\mu&=i\frac{\tilde g_V}{\sqrt{2}}
\begin{pmatrix}\tfrac{\rho^0_\mu+\omega_\mu}{\sqrt{2}}&\rho^+_\mu
&K^{*+}_\mu~\\
\rho^-_\mu&\tfrac{-\rho^0_\mu+\omega_\mu}{\sqrt{2}}&K^{*0}_\mu\\
K^{*-}_\mu&\bar K^{*0}_\mu&\phi_\mu\end{pmatrix}~,
\end{align}
where $\eta_8$ and $\eta_0$ contribute to $\eta - \eta^\prime$
mixing as in \cite{PDG} with
$\theta_P=-20\pm 5^0$.
The heavy pseudoscalar $D_a$ and vector $D^*_a$ fields of flavour $c\bar
q_a$
are incorporated in
\begin{equation}
H_a=\tfrac{1}{2}(1+\!\!\not{\!
v})[-D^v_a~\gamma_5+D^{*v}_{a\mu}~\gamma^\mu]~,\qquad \bar H_a=\gamma^0
H_a\gamma^0~.
\end{equation}

Above, $f=132$ MeV is the pseudoscalar decay constant and $\tilde g_V=5.8$
is the
$VPP$ coupling \cite{casalbuoni,hidden}. We fix $a=2$ assuming the exact
vector
meson dominance, when the  light pseudoscalars interact with the photon only
through the vector mesons \cite{casalbuoni,hidden,thesis}.
We shall use $g=0.59\pm 0.06$, obtained by CLEO from the measurement of the widths $D^{*+}\to D^0\pi^+$ and $D^{*+}\to D^+\pi^0$ \cite{CLEO.g}.
The
parameter $\kappa$ will eventually turn out to be multiplied by a small factor
$m_P^2$ in
the $D\to Pl^+l^-$ amplitudes  and its
contribution is negligible.

 The bosonized weak current coming from the light quarks is obtained by gauging
(\ref{lag})
\begin{equation}
\label{current.light}
\bar q_a\gamma^\mu(1-\gamma_5)q_b\simeq \bigl(if^2\xi [{\cal
A}^{\mu}+a({\cal
V}-\rho)^\mu]\xi^\dagger\bigr)_{ba}~~.
\end{equation}
 The weak current
${\bar q}_a\gamma^\mu(1-\gamma_5)c$ transforms under chiral $SU(3)_L\times
SU(3)_R$ transformation as  $({\bar
3}_L,1_R)$ and it is  linear in the
heavy meson fields $D^a$ and $D^{*a}_\mu$ \cite{BFO1,thesis}
\begin{align}
\label{current.heavy}
{\bar q}_a\gamma^\mu(1-\gamma^5)c &\simeq \tfrac{1}{2} i f_D\sqrt{m_D}  ~
 Tr
[\gamma^{\mu}
(1 - \gamma_{5})H_{b}\xi_{ba}^{\dag}]\\
&+ \alpha_{1}  Tr [\gamma_{5} H_{b} ({\rho}^{\mu}
- {\cal V}^{\mu})_{bc} \xi_{ca}^{\dag}]
+\alpha_{2} Tr[\gamma^{\mu}\gamma_{5} H_{b} v_{\alpha}
({\rho}^{\alpha}-{\cal V}^{\alpha})_{bc}\xi_{ca}^{\dag}]+...\nonumber
\end{align}
This  current  is the most general one at the leading order in the heavy
quark
 and next-to-leading order in the chiral expansion. The parameters
$\alpha_1$
and $\alpha_2$ are determined
 from  experimental data on $Br$,
$\Gamma_L/\Gamma_T$ and $\Gamma_+/\Gamma_-$ of the decay $D^+\to \bar
K^{*0}e^+\nu_e$
\cite{PDG}. Among the eight sets of solutions for three parameters
\cite{BFO1},
we use the set $\alpha_1=0.14\pm 0.01$ GeV$^{1/2}$ and $\alpha_2=0.10\pm
0.03$ GeV$^{1/2}$ which agrees with the measured form factors.

We shall calculate a larger group of
$D \to P l^+ l^-$ decays, rather than only those related
to $c \to u l^+ l^-$ transition. The list of decays considered is given in
Table \ref{tab2}.
The Feynman diagrams for the long distance contributions to $D\to Pl^+l^-$
within our framework are given in Fig. \ref{fig2}.  
The
Lagrangian (\ref{eff}) contains a product of two left handed quark currents,
each
denoted by
a dot in a box.  We organize different diagrams according to the
factorization
of the non-leptonic effective Lagrangian (\ref{eff}):

\begin{figure}[h]
\input{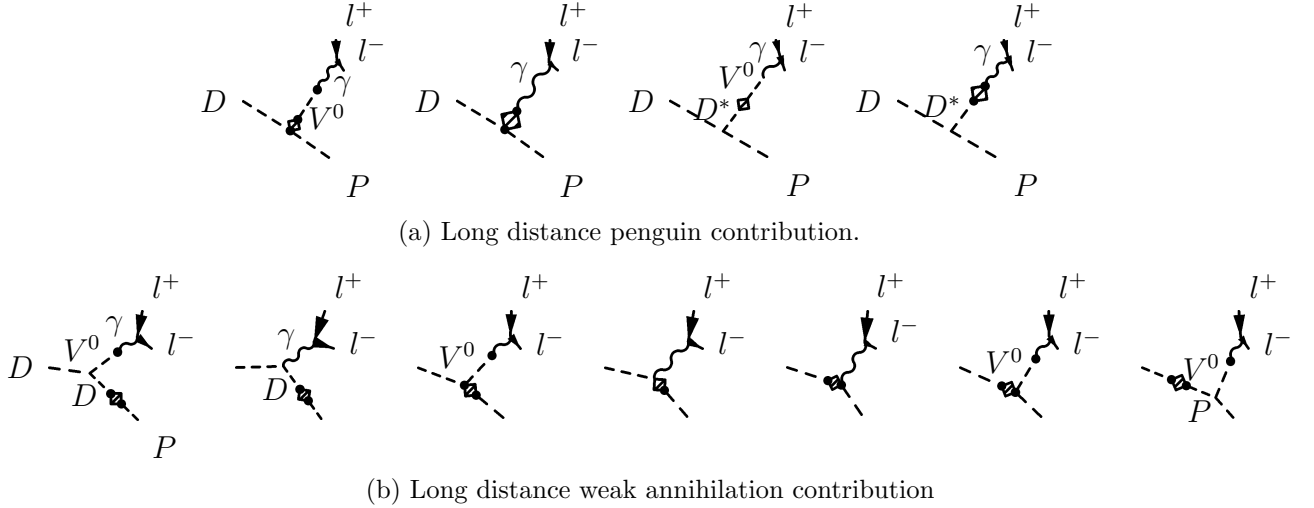}
\caption{Long distance contributions to  $D\to Pl^+l^-$ decays. The vector meson $V^0$ denotes $\rho^0$, $\omega$ or $\phi$. 
The box denotes the action of the nonleptonic effective Lagrangian (\ref{eff}). The box contains two dots each denoting a weak current in the Lagrangian (\ref{eff}).
}
\label{fig2}
\end{figure}

\begin{itemize}
\item
The {\bf long distance penguin} contribution \cite{sfps} in Fig. \ref{fig2}a is induced by 
$[\bar
s\gamma_\mu
s-\bar d\gamma_\mu d]~ u\gamma^{\mu}(1-\gamma_5)c$.

\item
The {\bf long distance weak annihilation} in Fig. \ref{fig2}b is induced by a product
of the
weak
currents, where  one  current has the flavour of the initial $D$ meson,
while
the other has the flavour of the final $P$ meson.
Vector resonances do not enter as intermediate states $R$ in the weak
transition
$D\to R$ followed by $R\to P\gamma^*$ or $D\to R \gamma^*$ followed by weak
transition $R\to P$, since parity is conserved in $D\to P\gamma^*$
process.

\end{itemize}

The Lagrangian (\ref{lag}) and the weak currents (\ref{current.light}),
(\ref{current.heavy}) are invariant under the electro-magnetic gauge
transformation and automatically lead to the gauge invariant
amplitude of
the form (\ref{1}). This is due to the fact that the vector field
$\rho_\mu$ and
the vector current ${\cal V}_\mu=ieQA_\mu+\tfrac{1}{2}(\xi^\dagger
\partial_\mu\xi+\xi \partial_\mu\xi^\dagger)$ always appear in the gauge
invariant combination ${\cal V}_\mu-\rho_\mu$ and the resonant and nonresonant diagrams in Fig. \ref{fig2} come in pairs.

We incorporate $SU(3)$ symmetry breaking by using the physical masses,
widths
and decay constants, given in Tables \ref{tab.const} and \ref{tab.excited} of  Appendix with the following definition
\begin{equation}
\label{def.decay}
 \langle 0|j^{\mu}|P\rangle =if_P p^{\mu},\quad \langle 0
|j^{\mu}|D\rangle
=-if_D p^{\mu},\quad \langle
0|j^\mu|V\rangle=g_V\epsilon^\mu,\quad
\langle 0|j^\mu|D^*\rangle=if_{D^*}m_{D^*}\epsilon^\mu
\end{equation}
and properly normalized $j^\mu=\bar q_1\gamma^\mu(1-\gamma_5)q_2$. The assumptions for extrapolating the amplitudes away from  where the chiral and heavy quark symmetries are good, are discussed in Appendix.
The amplitudes for the diagrams in Fig. \ref{fig2} 
 are given by Eq. (\ref{ald}).

\begin{table}[!htb]
\begin{center}
\begin{tabular}{|c|c c|c c|}
\hline
$D\to P l^+l^-$ & ${Br_{SM}^{SD}}$  
&${Br_{SM}\simeq Br^{LD}}$&${Br^{exp}}$&${Br^{exp}}$\\

 & $l=\mu,~e$&  $l=\mu,~e$&$l=e$&$l=\mu$\\
\hline 
$ D^0 \to {\bar K}^{0} l^+l^-$ & $0$&$4.3\times
10^{-7}$&$<1.1\times 10^{-4}$&$<2.6\times 10^{-4}$\\

 $ D_s^+ \to \pi^+ l^+l^-$ & $0$&$6.1\times 10^{-6}$&$<2.7\times 10^{-4}$&$<1.4\times 10^{-4}$\\
\hline
$ D^0 \to \pi^{0}l^+l^-$ & $1.9\times 10^{-9}$&$2.1\times 10^{-7}$&$<4.5\times 10^{-5}$&$<1.8\times 10^{-4}$\\

$ D^0 \to \eta l^+l^-$ & $2.5\times 10^{-10}$ &$4.9\times 10^{-8}$&$<1.1\times 10^{-4}$&$<5.3\times 10^{-4}$\\

$ D^0 \to \eta^\prime l^+l^-$ & $9.7\times 10^{-12}$ &$2.4\times 10^{-10}$&$<1.1\times 10^{-4}$&$<5.3\times 10^{-4}$\\

$ D^+ \to \pi^+ l^+l^-$ & $9.4\times 10^{-9}$&$1.0\times 10^{-6}$&$<5.2\times 10^{-5}$&$<7.8\times 10^{-6}$\\

$ D_s^+ \to K^{+ } l^+l^-$ & $9.0\times 10^{-10}$&$4.3\times 10^{-8}$&$<1.6\times 10^{-3}$&$<1.4\times 10^{-4}$\\
\hline

$ D^+ \to K^{+} l^+l^-$ & $0$&$7.1\times
10^{-9}$&$<2.0\times 10^{-4}$&$<8.1\times 10^{-6}$\\

$ D^0 \to K^{0} l^+l^-$ &$0$&$1.1\times 10^{-9}$&&\\
\hline
\end{tabular}
\caption{The branching ratios for nine $D\to Pl^+l^-$ decays in the
standard
model. The short distance contributions, induced by the $c\to ul^+l^-$
transition, are given in column 2 and are small. The total branching ratio
is
therefore dominated by the long distance contribution and is given in
column 3. The experimental upper
bounds
are given in the last two columns \cite{PDG,pll.exp,pll.focus}: the E791
analysis \cite{pll.exp} considers $D^+$ and $D^+_s$ decays,  
while the new analysis of FOCUS \cite{pll.focus} considers only $D^+$ 
decays. 
The MSSM has insignificant effect on the total rates of  $D\to Pl^+l^-$ 
decays.  }
\label{tab2}
\end{center}
\end{table}

\section{The results}

The allowed kinematical region for the di-lepton mass $m_{ll}$ in the $D\to
Pl^+l^-$ decay is $m_{ll}=[2m_l,m_D-m_P]$.
The long distance contribution has resonant shape with poles at
$m_{ll}=m_{\rho^0},m_\omega,m_\phi$. There is no  pole at $m_{ll}=0$
since the
decay $D\to P\gamma$ is forbidden. The short distance contribution is rather
flat.
The spectra of $D\to Pe^+e^-$ and $D\to P\mu^+\mu^-$ decays  in terms of
$m_{ll}$ are practically identical. The difference in their rates  due
to  the
kinematical region $m_{ll}=[2m_e,2m_\mu]$ is small  and we do not
consider them separately.  The predicted branching ratios for nine decays
in
the standard model are given in Table \ref{tab2} together with the
available
experimental data \cite{pll.exp,pll.focus,PDG}. The short distance contribution, as
predicted by the standard model, is given in the second column and is small. The total branching ratio
is
therefore dominated by the long distance contribution and is given in
column 3.

\begin{figure}[!htb]
\begin{center}
\includegraphics[scale=.35]{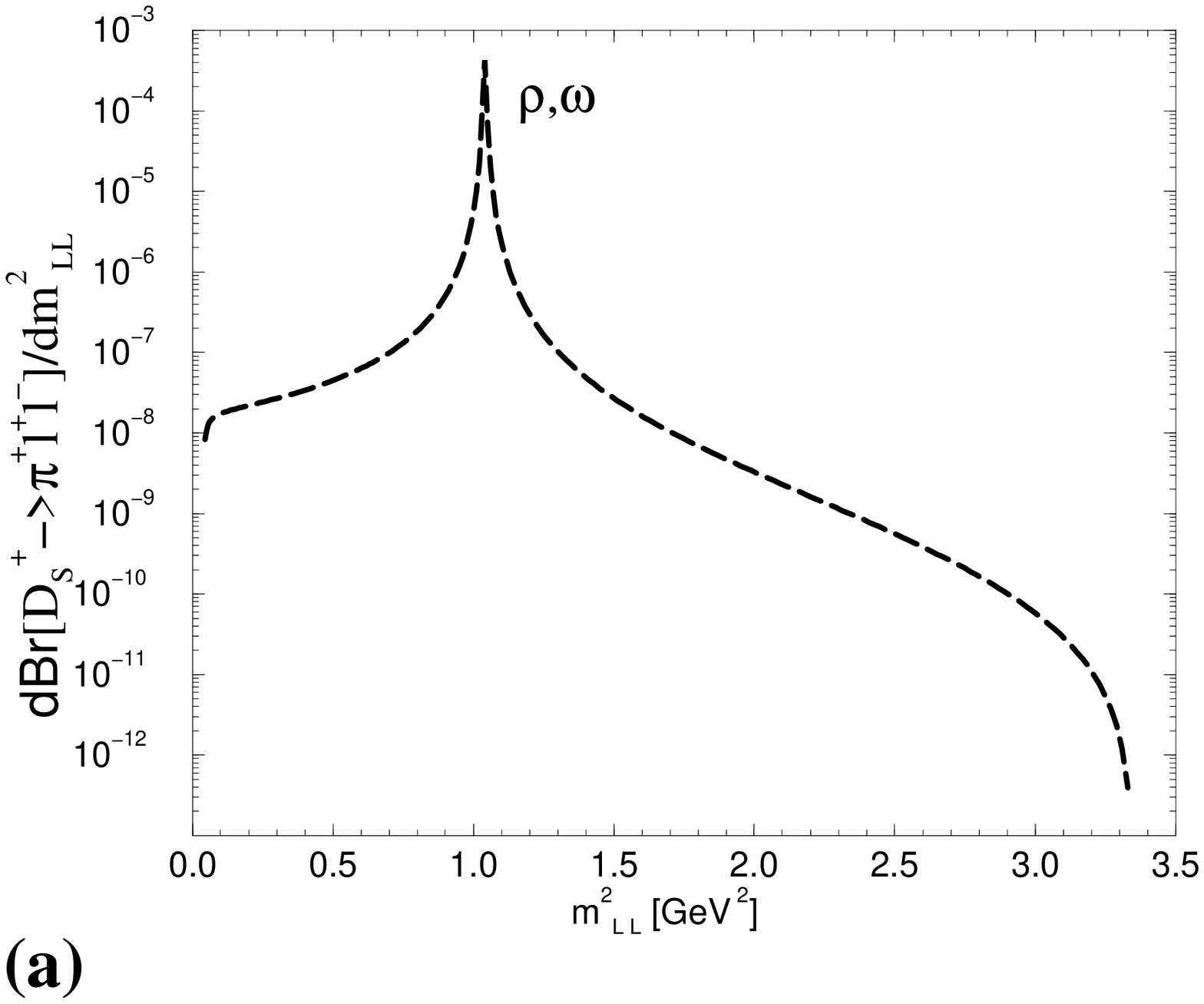}
\quad
\includegraphics[scale=.35]{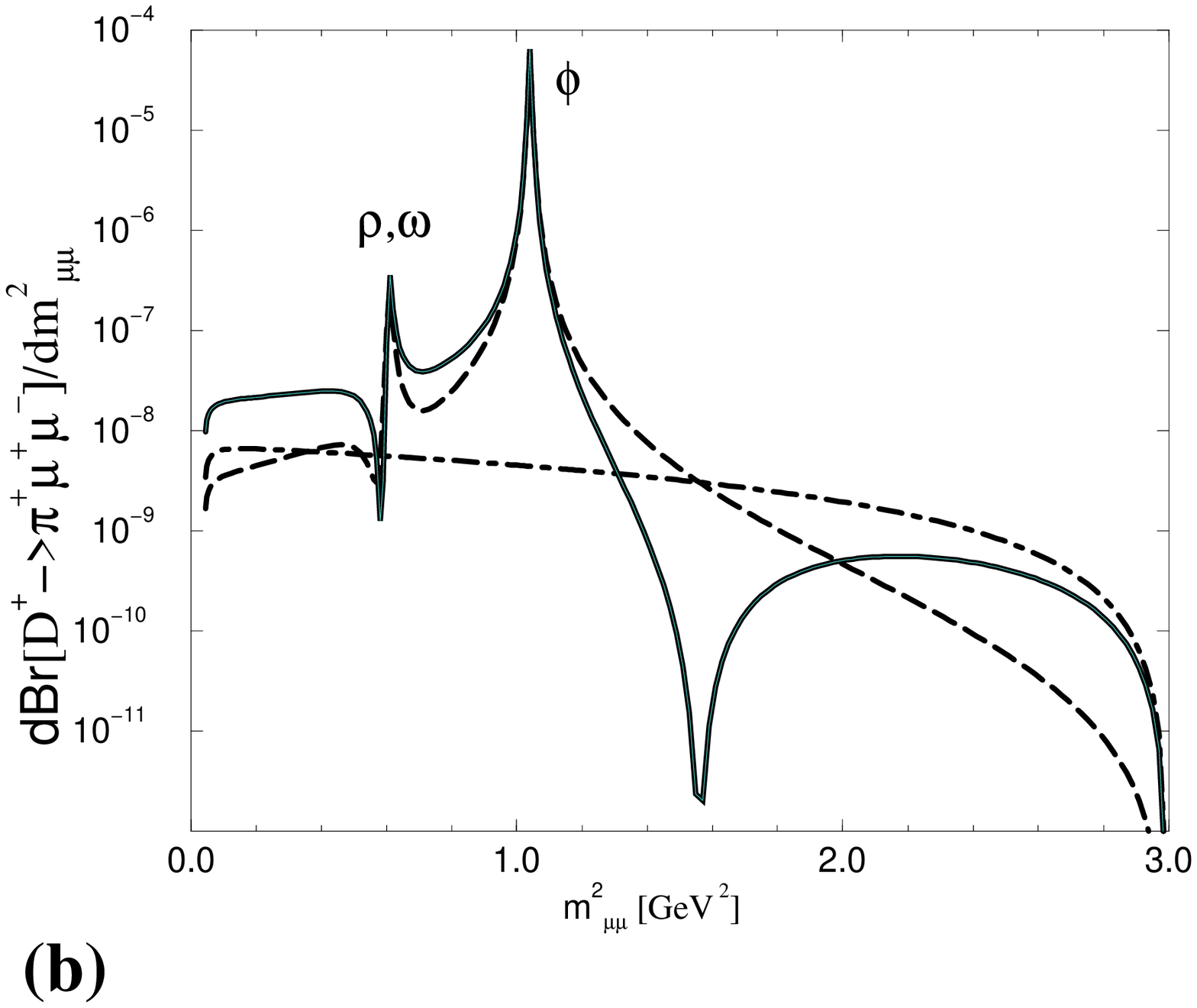}
 \caption
{The differential branching ratios $dBr/dm^2_{ll}$ as a function
of the
invariant di-lepton mass $m_{ll}^2$ for the  Cabibbo allowed decay 
$D_s^+\to \pi^+l^+l^-$ (a) and Cabibbo suppressed decay  
$D^+\to \pi^+l^+l^-$ (b).
The dashed
line denotes the long distance contribution, the dot-dashed line
denotes
the $c\to ul^+l^-$ induced short distance contribution, while the solid line denote the total standard model prediction. The $D_s^+\to \pi^+l^+l^-$ arises only via the long distance contribution. }
\label{fig4}
\end{center}
\end{figure}

The differential branching ratio $dBr/dm_{ll}^2$ for the Cabibbo allowed decay $D_s^+\to \pi^+l^+l^-$, which arises only via the  weak annihilation, is presented in
Fig. \ref{fig4}a.  In Fig. \ref{fig4}b, we present  the Cabibbo suppressed decay 
$D\to \pi l^+l^-$, in which the
kinematical upper bound on di-lepton mass $m_{ll}^{max}=m_D-m_P$ is  the
highest.
The dashed  and dot-dashed lines denote the long and short distance parts of the rate in SM, respectively, while the solid lines denote the total rate. The long distance contribution decreases  in the kinematical
region above the resonance $\phi$ and the short distance contribution
becomes
dominant. Thus, the decays $D^{+,0}\to \pi^{+,0} l^+l^-$ at high 
$m_{ll}$  might
present a
unique
opportunity to probe the flavour changing neutral transition $c \to
ul^+l^-$ in
the future. As the pion is the lightest hadron state, this interesting
kinematical region is not present in other  $D\to Xl^+l^-$ decays.

\begin{figure}[!htb]
\begin{center}
\includegraphics[scale=.35]{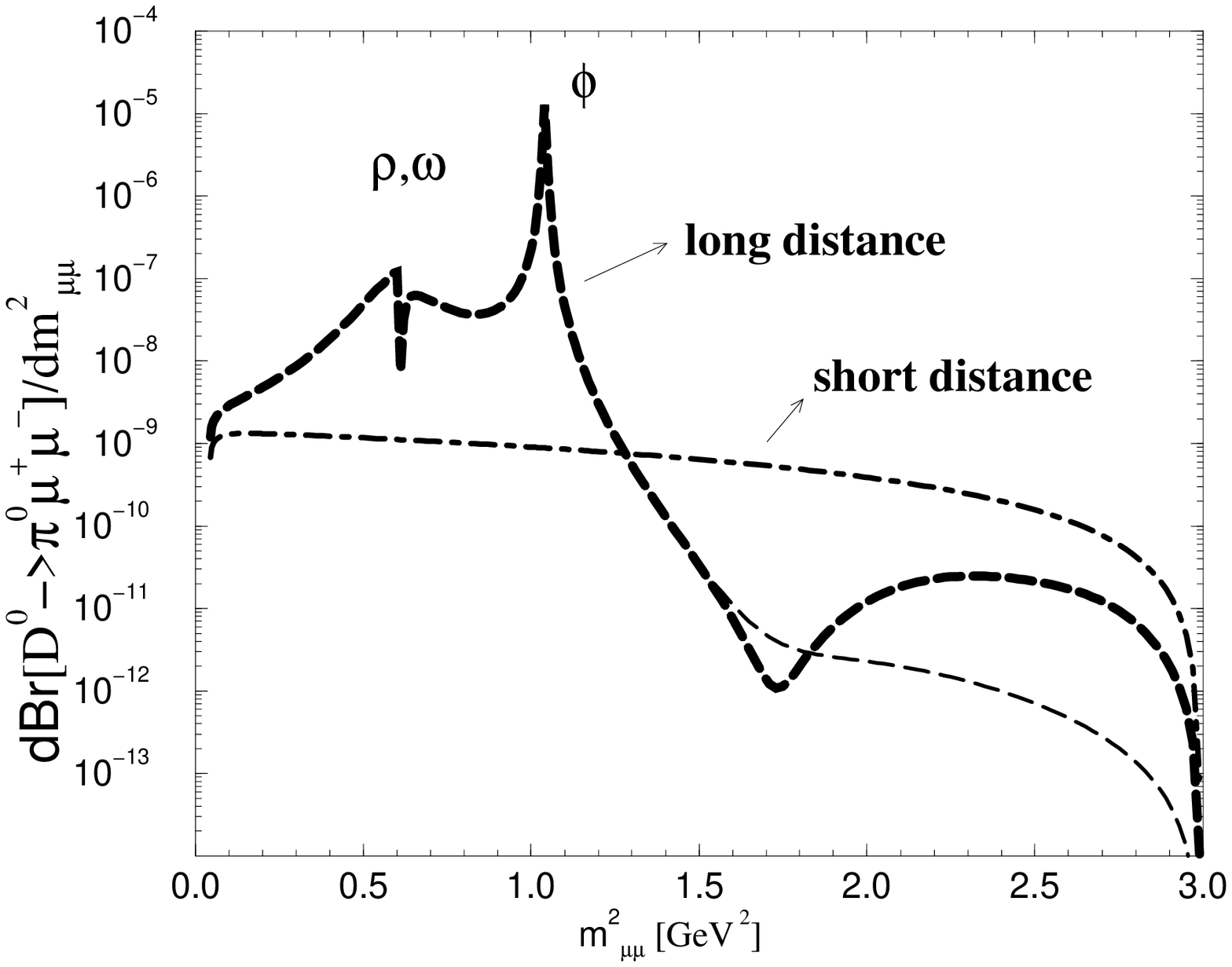}
\quad
\includegraphics[scale=.35]{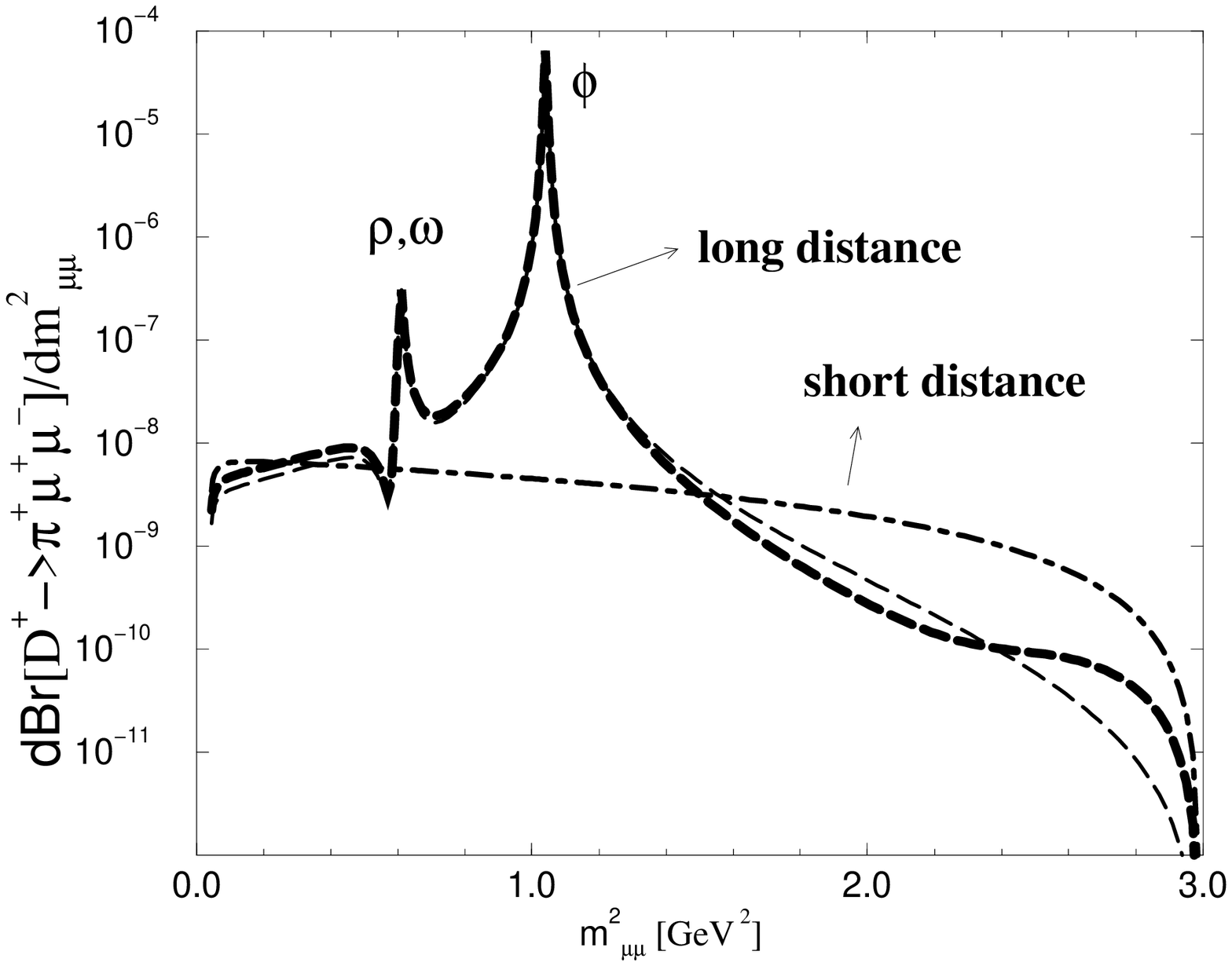}
\caption
{The differential branching ratio for  $D\to \pi \mu^+\mu^-$
decays.
The thick dashed lines present the long distance contribution incorporating the ground state and the excited vector mesons. The thin dashed lines present the long distance contributions due only to the ground vector mesons.   The short distance contribution, denoted by dot-dashed line, is dominant in the kinematical region of high $m_{ll}$, in spite of the excited vector resonances. }
\label{fig5}
\end{center}

\end{figure}

The differential distribution for $D^{+} \to \pi^{+} l^+ l^-$,
given in Fig. \ref{fig4}, indicates that the high  di-lepton mass region might
give an opportunity for detecting $c \to u l^+ l^-$. Before making a definite
statement on such possibility, we should
 examine this kinematical region of high di-lepton mass in $D\to \pi
l^+l^-$ decays more closely. For instance, in this region the excited states of
the vector
mesons $\rho$, $\omega$ and $\phi$ may become important.
We attempt  a  rough estimate of the additional long
distance
contribution arising from the first radial excited states   $\rho_1$,
$\omega_1$
and $\phi_1$  ($^3S_1$) and first orbital excited states $\rho_2$,
$\omega_2$
and $\phi_2$ ($^3D_1$). The knowledge of their masses, decay widths and
couplings to other particles is poor at present. We use the measured
masses and
widths,  taken from \cite{PDG,CDquark} and compiled in Table
\ref{tab.excited}.
Due to the lack of the experimental data on the leptonic decay widths
\cite{CDquark}, we use the magnitudes of the decay constants $g_V$  as
predicted
by the quark model in \cite{IG}\footnote{The decay constant  $f_V$,
defined in
\cite{IG}, is related to $g_V$, defined in (\ref{def.decay}), by:
$f_\rho\to \sqrt{2} m_\rho f_\rho$, $f_\omega\to 3\sqrt{2} m_\omega f_\omega$ and
$f_\phi\to -3 m_\phi f_\phi$.} and compiled in Table \ref{tab.excited}. At
the same
time, we assume that the excited vector mesons couple to the charmed mesons
with the same couplings as the corresponding ground state vector mesons
$\rho$,
$\omega$ and $\phi$.
In this case, the corresponding amplitudes (\ref{ald}) are obtained by replacing  the
coefficients $N_1$ and $M_1$  by the expressions given in
(\ref{replacement}).
The differential branching ratios for  $D\to \pi \mu^+\mu^-$ decays are
given in
Fig. \ref{fig5}. The thick and thin dashed lines denote the long distance contributions with and without excited vector mesons, respectively. The short distance contribution, denoted by dot-dashed line, is still dominant in the kinematical region of high $m_{ll}$ in spite of the excited vector resonances. 

\begin{figure}[!htb]
\begin{center}
\includegraphics[scale=.35]{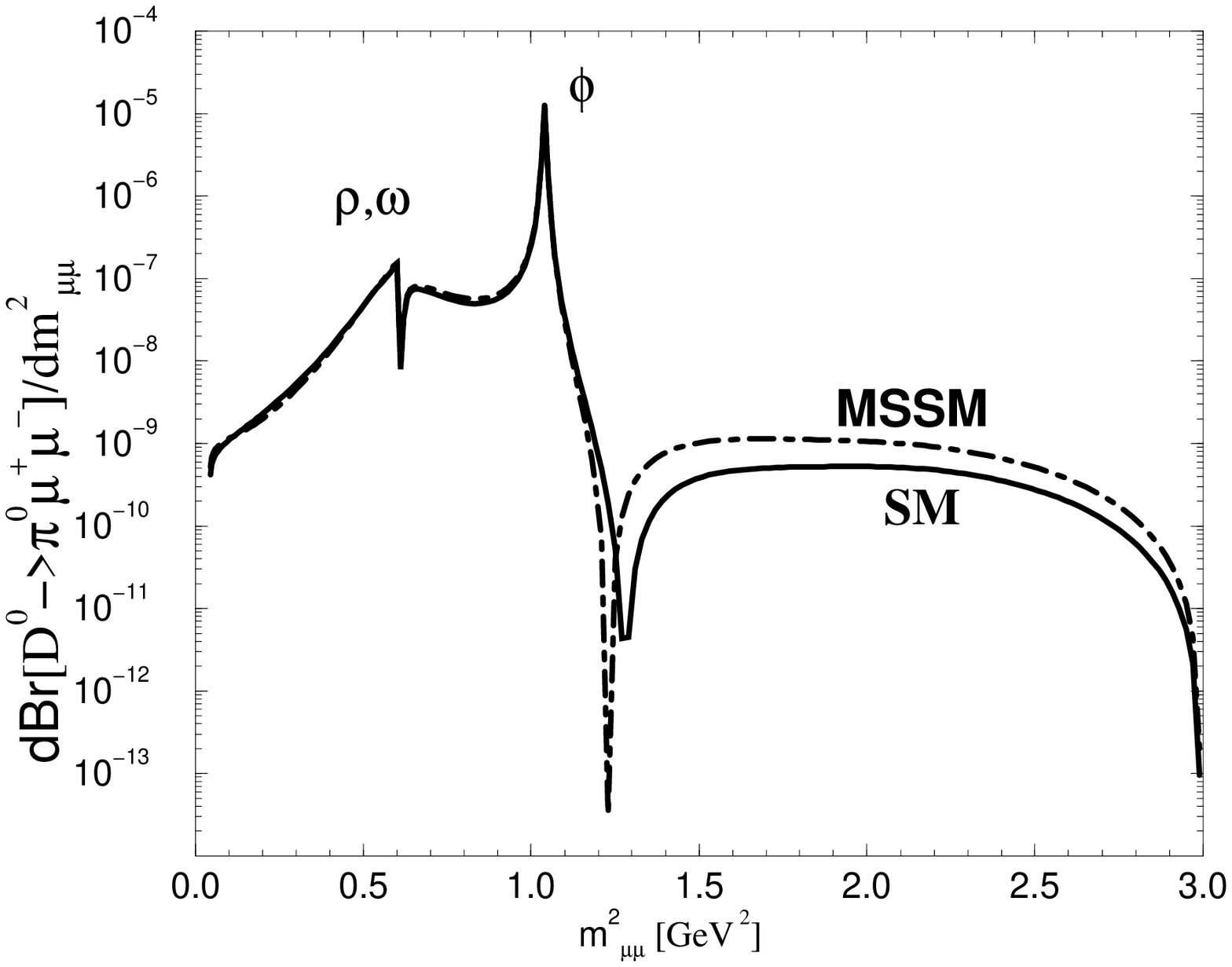}
\quad
\includegraphics[scale=.35]{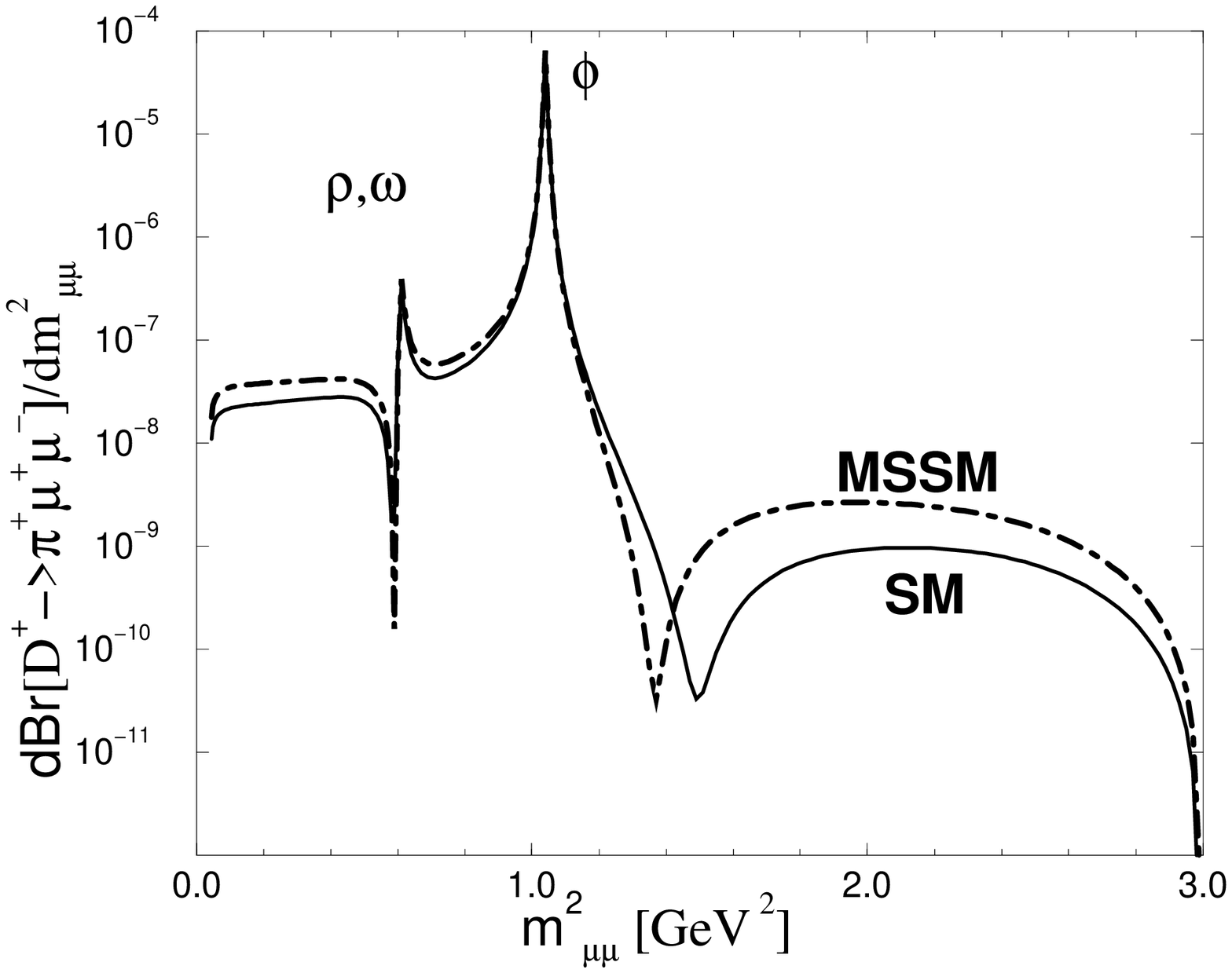}
 \caption
{The biggest possible enhancement of $D\to \pi\mu^+\mu^-$  rates within the
general
MSSM, discussed in section 2.1, is denoted by the dot-dashed lines. The solid lines represent the standard model
predictions. The effect of supersymmetry is screened by the uncertainties present in the determination of the long distance contributions and is probably too small to be observed.}
\label{fig6}
\end{center}
\end{figure}

The possible enhancement within the general MSSM, discussed in section
2.1, is
presented in Fig. \ref{fig6} and is probably too small to be observed in
any
$D\to Pl^+l^-$ decay. The solid lines represent the standard model
prediction
for the  $D\to
\pi l^+l^-$ branching ratios. The dot-dashed lines represent the best enhancement in the
general MSSM and indicate that the $D\to Pl^+l^-$ rates are rather
insensitive
to the large supersymmetric enhancement of  $c_7$. The value of $c_7$ is
manifested in  $c\to ul^+l^-$ at small $m_{ll}$ (see Eq. (\ref{br}) and Fig.
\ref{fig1}), while it's effect is suppressed in $D\to Pl^+l^-$ decays do to the factor $q^2$ in the general expression for the $D\to P\gamma^*$ amplitude (\ref{1}).

\section{Conclusions}

We have presented the first predictions for  rare charm meson decays $D\to Pl^+l^-$ with $P=\pi,K,\eta,\eta^\prime$ in all nine possible channels; the previous analysis \cite{singer} has considered only the $D\to \pi l^+l^-$ channel. The long distance contributions are found to dominate over the short distance contributions, which are induced by $c\to ul^+l^-$ in the Cabibbo-suppressed decays. We have used the theoretical framework of Heavy Meson Chiral Lagrangian with the recently determined value of the strong coupling $g$ from the measurement of $D^*\to D\pi$ width. Our predictions are compiled in Table \ref{tab2}. The decay $D^+_s\to \pi^+l^+l^-$ is predicted at the highest branching ratio of $6\times 10^{-6}$. The best chances of the experimental discovery are expected for $D^+\to \pi^+l^+l^-$, which is predicted at  $1\times 10^{-6}$ and has the upper bound  $8\times 10^{-6}$ \cite{pll.focus} at present. The limits on  $D^0$ and $D^+$ modes at the level $10^{-6}$ are expected from CLEO-c and B-factories, while the limits on $D_s^+$ modes are expected to be an order of magnitude  milder \cite{CLEOc}.

The only possibility to look for $c\to ul^+l^-$ transition is represented by $D\to \pi l^+l^-$ decays in the kinematical region of $m_{ll}$  above the resonance $\phi$, where the long distance contribution is reduced (see Fig. \ref{fig5}).

We have explored the sensitivity of the $c\to ul^+l^-$ within two scenarios of physics beyond SM. The effect due the exchange of the flavour changing Higgs in Two Higgs Doublet model is found to be negligible. The general Minimal Supersymmetric Standard model can enhance the  $c\to u\mu^+\mu^-$ rate by up to a factor of three (see Table \ref{tab1}). This effect is due to the large supersymmetric enhancement of  $c_7$  and is seizable at small $m_{ll}$ in   $c\to ul^+l^-$, but it is unfortunately very small  in the hadronic process $D\to Pl^+l^-$ as the decay $D\to P\gamma$ is forbidden (see Fig. \ref{fig6}).  

The kinematics of the processes $D\to Vl^+l^-$ would be more favorable to probe the possible supersymmetric enhancement at small $m_{ll}$, but the  long distance contributions in these channels  are even more disturbing \cite{FPS.vll}. 
The large supersymmetric enhancement of the Willson coefficient $c_7$  is
manifested
in $c\to u\gamma$ decay and can enhance the standard model  rate $\sim
10^{-8}$
by up to two orders of magnitudes \cite{sasa,masiero.cugamma}. Such
enhancement
could be probed by observation of  $B_c\to B_u^*\gamma$ \cite{FPS.bc} or by
measuring the relative difference $Br(D^0\to \rho^0\gamma)-Br(D^0\to
\omega\gamma)$ \cite{FPSW}.

\section{Appendix}

The short distance part of the $D\to Pl^+l^-$ amplitude, induced by the transition $c\to ul^+l^-$, contains the following form factors
\begin{align}
\label{formsd}
 \langle P(p^\prime)|\bar
q\gamma_\mu(1-\gamma_5)c|D(p)\rangle&=(p+p^\prime)_\mu f_+(q^2)+(p-p^\prime)_\mu f_-(q^2)\nonumber\\
\langle P(p^\prime)|\bar
q\sigma_{\mu\nu}(1\pm\gamma_5)c|D(p)\rangle&=
is(q^2)\bigl[(p+p^\prime)_\mu q_\nu-q_\mu(p+p^\prime)_\nu\pm i \epsilon_{\mu\nu\lambda\sigma}(p+p^\prime)^\lambda q^\sigma\bigr]
\end{align}
defined using operators in (\ref{ope.lunghi}). The short distance amplitude is then given by
\begin{equation}
\label{asd}
{\cal A}^{SD}[D(p)\to
P(p-q)l^+l^-]=i~\tfrac{G_F}{\sqrt{2}}e^2V_{cs}^*V_{us}\biggl[-\frac{c_7+c_7^\prime}{2\pi^2}m_c s(q^2)-\frac{c_9}{4\pi^2}f_+(q^2)\biggr]\bar u(p_-)
\!{\not \!p}v(p_+)~,
\end{equation}
where we neglected the nearly vanishing $c_{10}$, $c_9^\prime$ and $c_{10}^\prime$ coefficients in SM (\ref{c0.sm}) and MSSM (\ref{c.susy}).
In the heavy quark limit, the form factor $s$ can be expressed in terms
of the
form factors $f_\pm$ at zero recoil \cite{IW}\footnote{This relation is not written correctly in \cite{IW} and is corrected in \cite{casalbuoni}.}  and we assume the relation
to be
valid for all $q^2$
\begin{equation}
\label{forms}
s(q^2)=\frac{1}{2m_D}\bigl[f_+(q^2)-f_-(q^2)\bigr]~.
\end{equation}
The semileptonic form factors $f_{\pm}$ in the Heavy Meson Chiral Lagrangian approach, extended by assuming the polar shape, are given by  \cite{wise,casalbuoni}\footnote{
Different form factors $f_\pm$ were used together with $g\simeq 0.27$ in \cite{thesis}. These form factors would overproduce the semileptonic decay rates for the value $g\simeq 0.59$ recently measured by CLEO \cite{CLEO.g}.  }
\begin{equation}
\label{formP}
f_{+}(q^2)=-f_-(q^2)=-K_{DP}~\frac{f_D}{2}~\biggl[g\frac{m_D-m_P}{m_P+m_{D^{\prime *}}-m_D}\biggr]~\frac{m_{D^{\prime *}}^2-q^2_{max}}{m_{D^{\prime *}}^2-q^2}~.
\end{equation}
with $K_{DP}$ given in Table \ref{tab.pll}.

\vspace{0.2cm}

The {\bf long distance} amplitude is given by the diagrams in
Fig. \ref{fig2}.  The long distance penguin diagrams in Fig. \ref{fig2}a are expressed in terms of the form factor $f_+$ (\ref{formP}). The weak annihilation contribution in Fig. \ref{fig2}b is determined by assuming that the vertices do not change significantly away from the
kinematical region, where the heavy quark and chiral symmetries are good. We expect this to be a reasonable approximation in $D$ meson decays.  
At the same time we use the full
heavy meson propagators $1/(p_D^2-m^2)$ instead of the HQET propagators
$1/(2mvk)$ \cite{BFO1}.   In the limit $m_P\ll m_D$, the bremsstrahlung-like diagrams  in Fig. \ref{fig2}b cancel exactly, as explained in detail in the Sections 3.3.3 and 5.5.1 of \cite{thesis}. Only the non-bremsstrahlung weak annihilation diagrams in Fig. \ref{fig2}a render the non-vanishing contribution. The  long distance amplitude is given by \cite{thesis}
\begin{align}
\label{ald}
{\cal A}^{LD}[&D(p)\to
P(p-q)l^+(p_+)l^-(p_-)]=i~\tfrac{G_F}{\sqrt{2}}~e^2~A^{LD}(q^2)~\bar u(p_-)
\!{\not \!p}
v(p_+)~,\\
A^{LD}(q^2)&=A^{LD}_{peng.}(q^2)+
A^{LD}_{{ annih.\atop bremsstrahlung}}(q^2)+A^{LD}_{{annih.\atop
non-brem.}}(q^2)~,\nonumber\\
A^{LD}_{peng.}(q^2)&=a_2V_{cs}^*V_{us}\frac{1}{q^2}f_+(q^2)N_1(q^2)~,\nonumber\\
  A^{LD}_{annih.\atop bremsstrahlung}(q^2)&\simeq 0~,\nonumber\\
A^{LD}_{annih.\atop
non-brem.}(q^2)&=f_{Cabb}^{(i)}\frac{1}{q^2}M_1^{(i)}(q^2)f_P\biggl[-f_D
\kappa \frac{m_P^2}{m_D^2 -m_P^2}
-\sqrt{m_D}\bigl(\alpha_1-\frac{m_D^2+m_P^2-q^2}{2m_D^2}\alpha_2
\bigr)\biggr]\frac{\tilde g_V}{\sqrt{2}}~\nonumber
\end{align}
with Cabibbo factors $f_{Cabb}^{(i)}$ and the coefficients $M_1(q^2)$ and
$K_{DP}^{(i)}$ as given in Table \ref{tab.pll}. 
The coefficient $N_1$ equals
\begin{align*}
N_1(q^2)&={g_{\rho}^2\over
q^2-m_{\rho}^2+i\Gamma_{\rho}m_{\rho}}-{g_{\omega}^2\over
3(q^2-m_{\omega}^2+i\Gamma_{\omega}m_{\omega})}-{2g_{\phi}^2\over
3(q^2-m_{\phi}^2+i\Gamma_{\phi}m_{\phi})}\\
&+{g_{\rho}^2\over m_{\rho}^2}-{g_{\omega}^2\over
3m_{\omega}^2}-{2g_{\phi}^2\over 3m_{\phi}^2}~,
\end{align*}
while the coefficients $M_1^{(i)}$ are given in terms of  $M^{D^0}_1$,
$M^{D^+}_1$
and $M^{D_s^+}$ in Table \ref{tab.pll}
\begin{eqnarray}
\label{m1}
M^{D^0}_1\!\!\!&=&\!\!\!{g_{\rho}\over
q^2-m_{\rho}^2+i\Gamma_{\rho}m_{\rho}}+{g_{\omega}\over
3(q^2-m_{\omega}^2+i\Gamma_{\omega}m_{\omega})}+{g_{\rho}\over
m_{\rho}^2}+{g_{\omega}\over 3m_{\omega}^2}~,
\nonumber\\
M^{D^+}_1\!\!\!&=&\!\!\!-{g_{\rho}\over
q^2-m_{\rho}^2+i\Gamma_{\rho}m_{\rho}}+{g_{\omega}\over
3(q^2-m_{\omega}^2+i\Gamma_{\omega}m_{\omega})}-{g_{\rho}\over
m_{\rho}^2}+{g_{\omega}\over 3m_{\omega}^2}~,
\nonumber\\
M^{D_s^+}_1\!\!\!&=&\!\!\!-{2g_{\phi}\over
3(q^2-m_{\phi}^2+i\Gamma_{\phi}m_{\phi})}-{2g_{\phi}\over 3m_{\phi}^2}~.
\end{eqnarray}
Note that  $N_1(0)=M_1(0)=0$ for $\Gamma(0)=0$ and there is no pole
arising from
to the photon propagator at $q^2=0$. The relative sign of the short and long
distance penguin amplitudes agrees with the Ref. \cite{PR}, which is based
on
assumption of quark-hadron duality.

\begin{table}[h]
\begin{center}
\begin{tabular}{|c|c||c|c|c|}
\hline
$i$ & $D\to P l^+l^-$ & $f_{Cabb}^{(i)}$  & $M^{(i)}$  & $K_{DP}^{(i)}$\\
\hline \hline
$1$ & $ D^0 \to {\bar K}^{0} l^+l^-$ &$ a_2V_{ud} V_{cs}^*$ &
$M_1^{D^0}$ &
$0$
\\
\hline
$2$ & $ D_s^+ \to \pi^+ l^+l^-$ & $a_1 V_{ud} V_{cs}^* $ & $M_1^{D_s^+}$ &
$0$  \\
\hline
\hline
$3$ & $ D^0 \to \pi^{0}l^+l^-$ &$ -a_2V_{ud} V_{cd}^*$ &
$-\tfrac{1}{\sqrt{2}}M_1^{D^0}$ & $\tfrac{1}{\sqrt{2}f_\pi}$ \\
\hline
$4$ & $ D^0 \to \eta l^+l^-$ &$ a_2 V_{ud}V_{cd}^*$ &
$-\sqrt{\tfrac{3}{2}}M_1^{D^0}\cos\theta_P$ &
$\tfrac{\cos\theta_P}{\sqrt{6}f}-\tfrac{\sin\theta_P}{\sqrt{3}f}$  \\
\hline
$5$ & $ D^0 \to \eta^\prime l^+l^-$ &$ a_2 V_{ud}V_{cd}^*$ &
$-\sqrt{\tfrac{3}{2}}M_1^{D^0}\sin\theta_P$ &
$\tfrac{\sin\theta_P}{\sqrt{6}f}+\tfrac{\cos\theta_P}{\sqrt{3}f}$  \\
\hline
$6$ & $ D^+ \to \pi^+ l^+l^-$ &$ -a_1 V_{ud} V_{cd}^* $ &
$M_1^{D^+}$ &
$\tfrac{1}{f_\pi}$  \\
\hline
$7$ & $ D_s^+ \to K^{+ }l^+l^-$ &$ a_1 V_{ud} V_{cd}^* $ &
$M_1^{D_s^+}$
 & $\tfrac{1}{f_K}$ \\
\hline
\hline
$8$ & $ D^+ \to K^{+} l^+l^-$ &$ -a_1 V_{us} V_{cd}^*$ & $M_1^{D^+}$ & $0$ \\
\hline
$ 9$ & $ D^0 \to K^{0} l^+l^-$ &$ -a_2 V_{us} V_{ud}^*$ &  $M_1^{D^0}$ &
$0$ \\
\hline
\end{tabular}
\caption{ The Cabibbo factors $f_{Cabb}^{(i)}$, the coefficients
$K_{DP}^{(i)}$
and the functions $M_1^{(i)}$ for nine $D\to Pl^+l^-$ amplitudes in (\ref{ald}).}
\label{tab.pll}
\end{center}
\end{table}

\vspace{0.1cm}

\begin{table}[h]
\begin{center}
\begin{tabular}{|c|c|c||c|c|c|}
\hline
$H$ & $m_H$ & $f_H$ & $P$ & $m_P$ & $f_P$   \\
&[GeV] & [GeV] & & [GeV] & [GeV] \\
\hline
$D$ & $1.87$ & $0.21 $ & $\pi$ & $0.14$ & $0.135$  \\
$D_s$ & $1.97$ & $0.24 $ & $K$ & $0.50$ & $0.16$ \\
$D^*$ & $2.01$ & $0.21 $ & $\eta$ & $0.55$ & $0.13$ \\
 &  &  & $\eta^\prime$ & $0.96$ & $0.11$ \\
\hline
\end{tabular}
\end{center}
\caption{The values of the meson masses, decay constants and decay widths
\cite{PDG}. The measured decay constants  $f_D$ and $f_{D^*}$ have sizable
uncertainties and the values are taken from lattice QCD results
\cite{f.lattice}. }
\label{tab.const}
\end{table}

\begin{table}[h]
\begin{center}
\begin{tabular}{|c|ccc|ccc|ccc|}
\hline
& $\rho$&$\omega$&$\phi$&$\rho_1$&$\omega_1$
&$\phi_1$&$\rho_2$&$\omega_2$&$\phi_2$\\
\hline
m[GeV]& 0.77&0.78&1.0&1.45&1.46&1.69&1.66&1.66&1.88\\
$\Gamma$[GeV]& 0.15&0.0084&0.0044&0.31&0.24&0.3&0.4&0.1&0.3\\
$g_V$[GeV$^2$]& 0.17&0.15&0.24&0.11&0.11&0.23&0.07&0.07&0.12\\
\hline
\end{tabular}
\caption{The masses, widths and decay constants of ground \cite{PDG} and
excited \cite{CDquark,IG} vector mesons.}
\label{tab.excited}
\end{center}
\end{table}

In order to account for the contributions of the excited vector mesons
$\rho_{1,2}$, $\omega_{1,2}$ and $\phi_{1,2}$, as described in the main
text,
the coefficients $N_1$ and $M_1$ are replaced in the formulas above
(\ref{ald}),
(\ref{m1}) by
\begin{align}
\label{replacement}
 N_1&\to N_1+\sum_{k=1}^{2}{g_{\rho_k }^2\over
q^2-m_{\rho_k}^2+i\Gamma_{\rho_k}m_{\rho_i}}-{g_{\omega_k}^2\over
3(q^2-m_{\omega_k}^2+i\Gamma_{\omega_k}m_{\omega_k})}-{2g_{\phi_k}^2\over
3(q^2-m_{\phi_k}^2+i\Gamma_{\phi_k}m_{\phi_k})}\nonumber\\
&\qquad +{g_{\rho_k}^2\over m_{\rho_k}^2}-{g_{\omega_k}^2\over
3m_{\omega_k}^2}-{2g_{\phi_k}^2\over 3m_{\phi_k}^2}~,\nonumber\\
M^{D^0}_1&\to M^{D^0}_1+\sum_{k=1}^{2}{g_{\rho_k}\over
q^2-m_{\rho_k}^2+i\Gamma_{\rho_k}m_{\rho_k}}+{g_{\omega_k}\over
3(q^2-m_{\omega_k}^2+i\Gamma_{\omega_k}m_{\omega_k})}+{g_{\rho_k}\over
m_{\rho_k}^2}+{g_{\omega_k}\over 3m_{\omega_k}^2}~,
\nonumber\\
M^{D^+}_1&\to M^{D^+}_1-\sum_{k=1}^{2}{g_{\rho_k}\over
q^2-m_{\rho_k}^2+i\Gamma_{\rho_k}m_{\rho_k}}+{g_{\omega_k}\over
3(q^2-m_{\omega_k}^2+i\Gamma_{\omega_k}m_{\omega_k})}-{g_{\rho_k}\over
m_{\rho_k}^2}+{g_{\omega_k}\over 3m_{\omega_k}^2}~,
\nonumber\\
M^{D_s^+}_1&\to M^{D_s^+}-\sum_{k=1}^{2}{2g_{\phi_k}\over
3(q^2-m_{\phi_k}^2+i\Gamma_{\phi_k}m_{\phi_k})}-{2g_{\phi_k}\over
3m_{\phi_k}^2}~.
\end{align}

\end{document}